\documentclass[prl,twocolumn,showpacs,amsmath,amssymb,superscriptaddress]{revtex4-1}
\usepackage{graphicx}
\usepackage{hyperref}
\usepackage{breqn}
\usepackage{xcolor}
\usepackage{color}
\usepackage{booktabs}
\usepackage{float}
\usepackage{longtable}
\usepackage{epsfig}

\makeatletter
\let\cat@comma@active\@empty
\makeatother
\begin{document}

\title{Giant and tunneling magnetoresistance effects from anisotropic and valley-dependent spin-momentum interactions in antiferromagnets}
\author{Libor \v{S}mejkal}
\affiliation{Institut f\"ur Physik, Johannes Gutenberg Universit\"at Mainz, D-55099 Mainz, Germany}
\affiliation{Institute of Physics, Czech Academy of Sciences, Cukrovarnick\'{a} 10, 162 00 Praha 6 Czech Republic}
\author{Anna Birk Hellenes}
\affiliation{Institut f\"ur Physik, Johannes Gutenberg Universit\"at Mainz, D-55099 Mainz, Germany}
\author{Rafael Gonz\'{a}lez-Hern\'{a}ndez }
\affiliation{Grup de Investigaci\'{o}n en F\'{i}sica Aplicada, Departamento de F\'{i}sica, Universidad del Norte, Barranquilla, Colombia}
\author{Jairo Sinova}
\affiliation{Institut f\"ur Physik, Johannes Gutenberg Universit\"at Mainz, D-55099 Mainz, Germany}
\affiliation{Institute of Physics, Czech Academy of Sciences, Cukrovarnick\'{a} 10, 162 00 Praha 6 Czech Republic}
\author{Tom\'a\v{s}  Jungwirth}
\affiliation{Institute of Physics, Czech Academy of Sciences, Cukrovarnick\'{a} 10, 162 00 Praha 6 Czech Republic}
\affiliation{School of Physics and Astronomy, University of Nottingham, Nottingham NG7 2RD, United Kingdom}

\date{\today}

\begin{abstract}
Giant or tunneling magnetoresistance are physical phenomena used for  reading  information in commercial spintronic devices. The effects rely on a conserved spin current passing between a reference and a sensing ferromagnetic electrode in a multilayer structure. Recently, we have proposed that these fundamental spintronic effects can be realized in collinear antiferromagnets with staggered spin-momentum exchange interaction, which generates conserved spin currents  in the absence of a net equilibrium magnetization. Here we elaborate on the proposal by presenting archetype model mechanisms for the antiferromagnetic giant and tunneling magnetoresistance effects. The models are based, respectively, on  anisotropic and valley-dependent forms of the non-relativistic staggered spin-momentum interaction. Using first principles calculations we link these model mechanisms to real antiferromagnetic materials and predict a $\sim$100\% scale for the effects. We point out that besides the GMR/TMR detection, our models directly imply the possibility of spin-transfer-torques excitation of the antiferromagnets. 
\end{abstract}

\maketitle
In the non-relativistic band structure of ferromagnets, the exchange interaction induces an energy gap between spin-up and spin-down states, making one spin state more populated and the other one less.  This results in different ohmic resistivities of the majority and minority spin channels, and in spin-dependent density of states. The former is  the basis of the giant magnetoresistance (GMR) effect in a trilayer stack comprising two ferromagnetic metal electrodes separated by a non-magnetic metal spacer \cite{Chappert2007}. The different densities of states of the majority and minority spins then govern the tunneling magnetoresistance (TMR) in a trilayer with a tunnel barrier between the ferromagnetic electrodes \cite{Chappert2007}. In both GMR and TMR, the resistance of the stack depends on whether the ferromagnetic electrodes are magnetized parallel or antiparallel, with typically the higher resistance state corresponding to the antiparallel configuration. The well conserved spin currents, enabling the GMR/TMR readout of the magnetization reversal,  can also facilitate efficient electrical switching of the magnetization via a spin-transfer torque (STT) \cite{Ralph2008}. The superior on-off characteristics of the devices with STT switching and GMR/TMR readout  allowed spintronics to revolutionize the magnetic storage and memory industry \cite{Chappert2007,Brataas2012}.

Traditionally, exchange-split bands have been considered to be excluded in antiferromagnets due to the alternating directions of atomic moments in the crystal \cite{Turov1965,Neel1971,Nunez2006,Surgers2014,Surgers2016,Ghimire2018}. A characteristic example are antiferromagnets with a symmetry combining time-reversal with space-inversion, which results in Kramers spin-degeneracy of electronic bands over the entire Brillouin zone \cite{Smejkal2016}. Considering such a model antiferromagnet, a STT mechanism was theoretically proposed more than a decade ago which differs fundamentally from STT in ferromagnets  \cite{Nunez2006}. It is based on transmitting a staggered spin polarization from one to the other antiferromagnet where the spin polarization and the antiferromagnetic orders in the electrodes are all commensurate \cite{Nunez2006}. This is a subtle, spin-coherent quantum-interference phenomenon relying on perfectly epitaxial commensurate multilayers \cite{Nunez2006,MacDonald2011,Saidaoui2014}. Similarly delicate were the proposed GMR and TMR  effects in these antiferromagnetic structures \cite{MacDonald2011} which might explain why a non-relativistic spintronics concept based on antiferromagnets with spin-degenerate bands has not been experimentally viable. 

Research in electrical detection and manipulation of the antiferromagnetic order turned, instead, to reading and writing principles based on relativistic spin-orbit coupling phenomena \cite{Shick2010,Park2011b,Marti2014,Zelezny2014,Wadley2016,Jungwirth2016,Wadley2018,Chen2014,Kubler2014,Nakatsuji2015,Nayak2016}. The resulting successful demonstrations of experimental memory devices, showcasing among others the insensitivity of antiferromagnets to magnetic fields  or their ultra-fast dynamics, prompted extensive fundamental and applied research interest in antiferromagnetic spintronics \cite{Jungwirth2016,Zelezny2018,Nemec2018,Gomonay2018,Smejkal2017b,Baltz2018,Song2018b,Siddiqui2020}. Until recently, however, the realization of antiferromagnetic counterparts of the non-relativistic GMR, TMR and STT phenomena, driven by robust conserved spin currents, has remained elusive. 

The breakthrough,  our present work builds upon, is the discovery of macroscopic time-reversal symmetry breaking in collinear antiferromagnets by a staggered spin-momentum exchange interaction \cite{Reichlova2020}. It opens the possibility of electrical detection of the antiferromagnetic N\'eel  vector reversal as already demonstrated, both theoretically and experimentally, by the antiferromagnetic Hall effect \cite{Smejkal2020,Feng2020a,Reichlova2020}. The staggered spin-momentum exchange interaction leads to spin splitting in the non-relativistic band structure of a magnitude comparable to ferromagnets while preserving a fully compensated antiferromagnetic state with zero net magnetization \cite{Smejkal2020,Ahn2019,Hayami2019,Yuan2020,Feng2020a,Reichlova2020,Gonzalez-Hernandez2020}.  In combination with the calculated conserved highly-polarized spin currents \cite{Gonzalez-Hernandez2020}, the staggered spin-momentum interaction has been identified as a mechanism leading to robust antiferromagnetic GMR, TMR and STT phenomena \cite{Reichlova2020}. 

 \begin{figure}[tb]
	\centering
	\includegraphics[width=.5\textwidth]{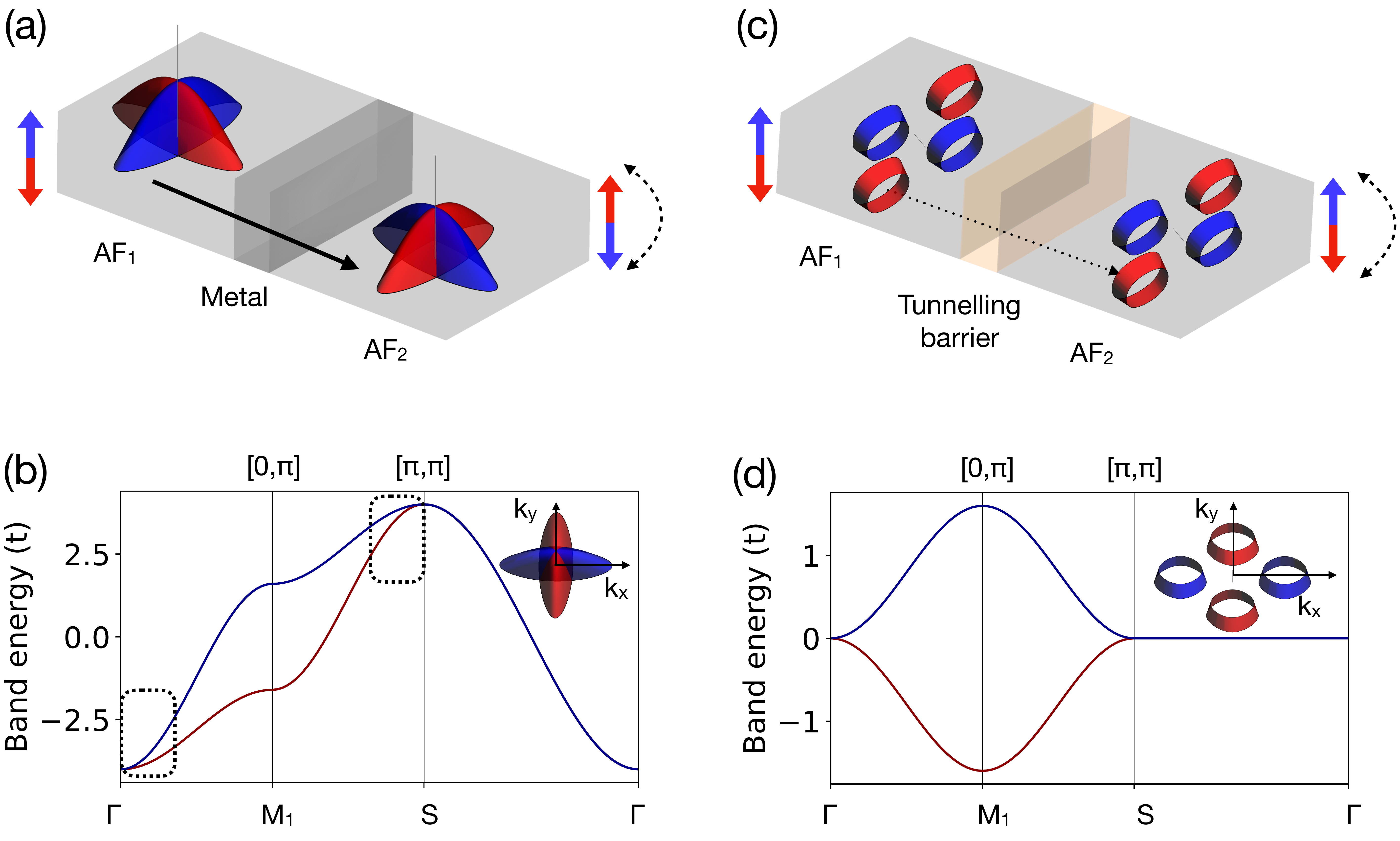} 
	\caption{\footnotesize Archetype antiferromagnetic GMR and TMR model mechanisms. Red and blue colors correspond to up and down spins. (a) GMR stack with a metallic spacer. As an example, we show the antiparallel configuration of the N\'eel vectors in the two antiferromagnetic electrodes AF$_1$ and AF$_2$. In the antiferromagnets, we show the energy band cuts highlighting the anisotropic staggered spin-momentum interaction around the $\boldsymbol\Gamma$ ({\bf S}) point in the Brillouin zone. The interfaces are oriented along one of the main axes of the elliptic bands.  (b) GMR model band dispersion. Dashed rectangles highlight the regions with the anisotropic staggered spin-momentum interaction. The inset corresponds to the region around the {\bf S}-point. (c) TMR stack with an insulating barrier.  As an example, we show the parallel configuration of the N\'eel vectors in the two antiferromagnetic electrodes.  In the antiferromagnets, we show the energy band cuts highlighting the valley-dependent staggered spin-momentum interaction around the {\bf M}$_1$ and {\bf M}$_2$ points in the Brillouin zone. (d) TMR model band dispersion with the inset highlighting isotropic regions around the {\bf M}$_1$ and {\bf M}$_2$ points with the staggered spin polarization.} 
	\label{fig1}
\end{figure}

In this paper we present archetype antiferromagnetic GMR and TMR models with staggered spin-momentum interaction. We then discuss the correspondence of these models to collinear antiferromagnetic phases of RuO$_2$ \cite{Smejkal2020,Ahn2019,Hayami2019,Feng2020a,Gonzalez-Hernandez2020} and Mn$_5$Si$_3$ \cite{Reichlova2020}. Using first principles calculations of spin-dependent conductivities and densities of states we  estimate the magnitudes of the  effects in RuO$_2$ and Mn$_5$Si$_3$. Finally we discuss the STT phenomenon in our   staggered spin-momentum interaction models.

The antiferromagnetic GMR and TMR models shown in Fig.~1 are based on a 2D tight-binding Hamiltonian with the kinetic nearest-neighbor hopping parametrized by $t$ and the staggered spin-momentum interaction parametrized by an exchange hopping $t_J$ \cite{Reichlova2020},
\begin{equation}
\mathcal{H}(\textbf{k})=2t\left( \cos {k_{x}}+ \cos {k_{y}} \right)\boldsymbol{1} + 2t_{J}\left( \cos {k_{x}}- \cos {k_{y}} \right) \boldsymbol\sigma\cdot \textbf{d}.
\label{Hamiltonian}
\end{equation}
Here $\textbf{d}$ is a unit vector along the N\'{e}el vector, $\boldsymbol{1}$ is the unit matrix and $\boldsymbol\sigma$ is the vector of Pauli spin matrices. 
The spin-up (+) and spin-down (-) energy bands are then given by,
\begin{equation}
E_{\pm}(\textbf{k})=2t\left( \cos {k_{x}}+ \cos {k_{y}} \right)\pm 2t_{J}\left( \cos {k_{x}}- \cos {k_{y}} \right).
\label{spectrum}
\end{equation} 

The antiferromagnetic GMR model shown in Fig.~1a,b is obtained by performing the ${\bf k}\cdot{\bf p}$ approximation around the $\boldsymbol\Gamma$-point, for which the energy spectrum can be written as 
\begin{equation}
E_{\pm}(\textbf{k})=4t+(t\pm t_J)k_x^2+(t\mp t_J)k_y^2,
\label{GMR-model}
\end{equation} 
and by taking, e.g., $t<0$ and $t_J=-0.4t$. The anisotropic spin-split bands, with equal net population of   spin-up and spin-down states, imply spin-dependent anisotropic conductivities, $\sigma_{+,xx}\neq\sigma_{-,xx}$, $\sigma_{+,yy}\neq\sigma_{-,yy}$, and $\sigma_{\pm,xx}=\sigma_{\mp,yy}$. Applying the current along, e.g.,  $x$-direction, the conductivity of the GMR stack will depend on whether the N\'eel vectors in the two antiferromagnets are parallel or antiparallel, in analogy to the ferromagnetic GMR. From the ratio of the conductivities of the spin-up and spin-down channels, $R_{\sigma}=\sigma_{+,xx}/\sigma_{-,xx}=\sigma_{+,xx}/\sigma_{+,yy}=\sigma_{-,yy}/\sigma_{-,xx}$, the antiferromagnetic GMR can be then estimated from the conventional current-in-plane GMR expression derived in ferromagnets \cite{Chappert2007}, 
\begin{equation}
{\rm GMR}=\frac{1}{4}(R_{\sigma}+\frac{1}{R_{\sigma}}-2).
\label{GMR}
\end{equation}
Below, in the section on first-principles calculations in RuO$_2$, we will use this expression to make quantitative estimates  in a real material realization of our antiferromagnetic GMR model (\ref{GMR-model}).

The TMR model in the antiferromagnet with the staggered spin-momentum interaction is shown in Fig.~1c,d. It is obtained from Eq.~(\ref{spectrum}) by performing the ${\bf k}\cdot{\bf p}$ approximation in {\bf M}$_1$ and {\bf M}$_2$ valleys. For $t=0$ we obtain, 
\begin{eqnarray}
E_{\pm}(\textbf{M}_1,\textbf{k})&=&\pm t_J(4-k^2), \nonumber \\
E_{\pm}(\textbf{M}_2,\textbf{k})&=&\mp t_J(4-k^2).
\label{TMR-model}
\end{eqnarray} 
The opposite spin splittings in the two valleys preserve the equal net population of   spin-up and spin-down states, while the  densities of states within a given valley become spin dependent, $n_+(\textbf{M}_1)\neq n_-(\textbf{M}_1)$,  $n_+(\textbf{M}_2)\neq n_-(\textbf{M}_2)$, and $n_{\pm}(\textbf{M}_1)=n_{\mp}(\textbf{M}_2)$. For tunneling which conserves the valley index, parallel and antiparallel configurations of the N\'eel vectors in the two antiferromagnets separated by the tunnel barrier will give different conductances, in analogy to ferromagnetic TMR.  We can then apply the Julli\`ere TMR formula \cite{Chappert2007} per valley,
\begin{equation}
{\rm TMR}=\frac{1}{2}(R_{n}+\frac{1}{R_{n}}-2),
\label{GMR}
\end{equation}
where the ratio of the spin-up and spin-down densities of states in the valley $R_{n}=n_+(\textbf{M}_1)/n_-(\textbf{M}_1)=n_+(\textbf{M}_1)/n_+(\textbf{M}_2)=n_-(\textbf{M}_2)/n_-(\textbf{M}_1)$. Again, we will use this expression below to make first principles calculation estimates of antiferromagnetic TMR in  RuO$_2$.

Before discussing the {\em ab initio} results, we show in Fig.~2 numerical conductivity calculations in a model structure with semi-infinite antiferromagnetic leads separated by a conductive spacer to represent GMR, or by an insulating spacer to represent TMR.  In both cases, the leads and the spacer (Fig.~2a) are modelled by the Hamiltonian (\ref{Hamiltonian}). For GMR, $t_J=t/2$ in the antiferromagnetic leads, while in the spacer $t$ is the same as in the leads and $t_J=0$. For the TMR calculations, we set  $t=0$ in the leads. In the tunneling spacer, $t_J=0$ and we set $t$ to 1/4 of $t_J$ in the leads, plus added an on-site energy in the spacer which is  $40$ times the hopping energy $t$ in the leads. The transport calculations were done using the Kwant package \cite{Groth2014a} (for more details see Supplementary information).

 \begin{figure}[tb]
	\centering
	\includegraphics[width=.5\textwidth]{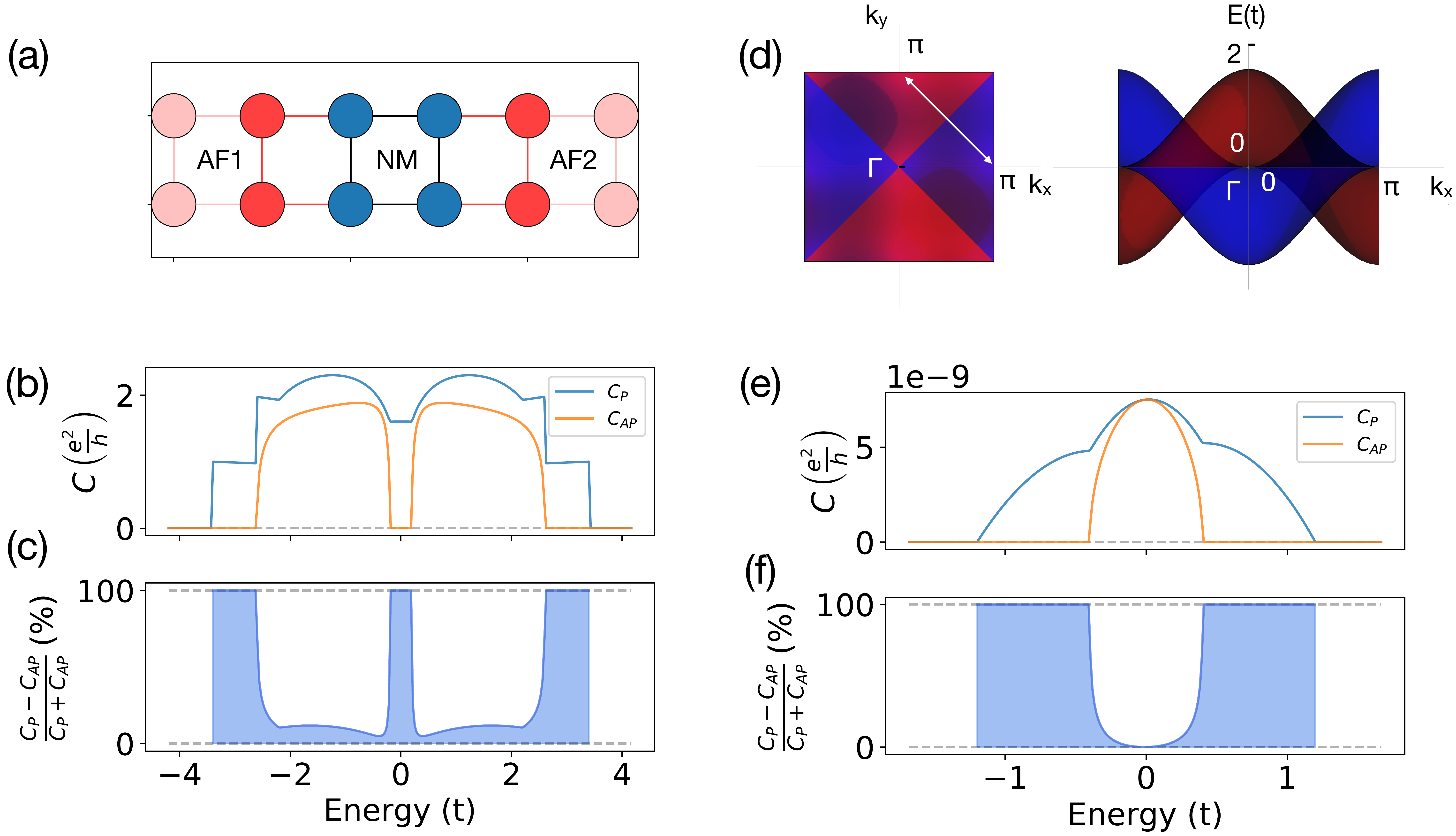} 
	\caption{\footnotesize Model transport calculations in antiferromagnetic GMR and TMR structures. (a) Schematics of the geometry of antiferromagnetic leads and the scattering region separating the leads, used in our calculations. (b) Conductances for parallel (P) and antiparallel (AP) configuraitons of the antiferromagnetic N\'eel vectors calculated for the GMR model with anisotropic staggered spin-momentum interaction (see text and Fig.~1a,b). (c) Corresponding relative difference of the P and AP conductances. Broad plateaux with large values correspond to regions near $\boldsymbol\Gamma$ and {\bf S} points (cf. Fig.~1(b)). (d) Top (left) and side (right) view of the spin polarised energy bands used in the tunneling magnetoresistance calculations. (e)  P and AP conductances in the tunneling stack calculated for the TMR model with valley-dependent  staggered spin-momentum interaction (see text and Fig.~1c,d). (f) Corresponding relative difference of the P and AP conductances. Plateaux with with large values correspond to the {\bf M}$_1$ ({\bf M}$_2$) valley (cf. Fig.~1(d)).} 
	\label{fig2}
\end{figure}

As anticipated in the above discussion of Fig.~1, we obtain different conductances for the parallel and antiparallel N\'eel vector configurations  in both the GMR and TMR variants of our model of staggered spin-momentum interaction in the antiferromagnetic leads (Figs.~2b,e). Also consistent with the qualitative conclusions based on Fig.~1, the difference between the parallel and antiparallel configurations is maximaized in the GMR structure when the Fermi energy crosses the bands near the $\boldsymbol\Gamma$-point (cf. Fig.~2c and Fig.~1b). In the TMR structure, on the other hand, the difference is maximized  at Fermi level positions where the  {\bf M}$_{1(2)}$ valleys dominate the tunneling transport (cf. Fig.~2f and Fig.~1d). This is again in agreement with the qualitative discussion of Fig.~1. 

We now turn to the first principles calculations in the collinear room-temperature antiferromagnet RuO$_2$. We calculated the electronic structure in the pseudopotential DFT code Vienna Ab initio Simulation Package (VASP) \cite{Kresse1996a}, within Perdew-Burke-Ernzerhof (PBE) + U + SOC (a spherically
invariant version of DFT + U), and we used an energy cutoff of 500~eV. We set the Hubbard $U$ to 1.6~eV. The lattice parametrs are $a = b = 4.5331$, $c = 3.1241$~\AA \, \cite{Smejkal2020}. Wannier functions were obtained  using the Wannier90 code \cite{Mostofi2014}. The longitudinal conductivity was calculated separately for each spin channel using Boltzamann equation with a $160^3$ crystal momentum mesh, and with the scattering rate of 6.6~meV corresponding to the experimental conductivity \cite{Feng2020a}.  (We have also confirmed the results by calculations using the WannierBerri code \cite{Tsirkin2021}, which we show in the Supplementary information.)

 \begin{figure}[tb]
	\centering
	\includegraphics[width=.45\textwidth]{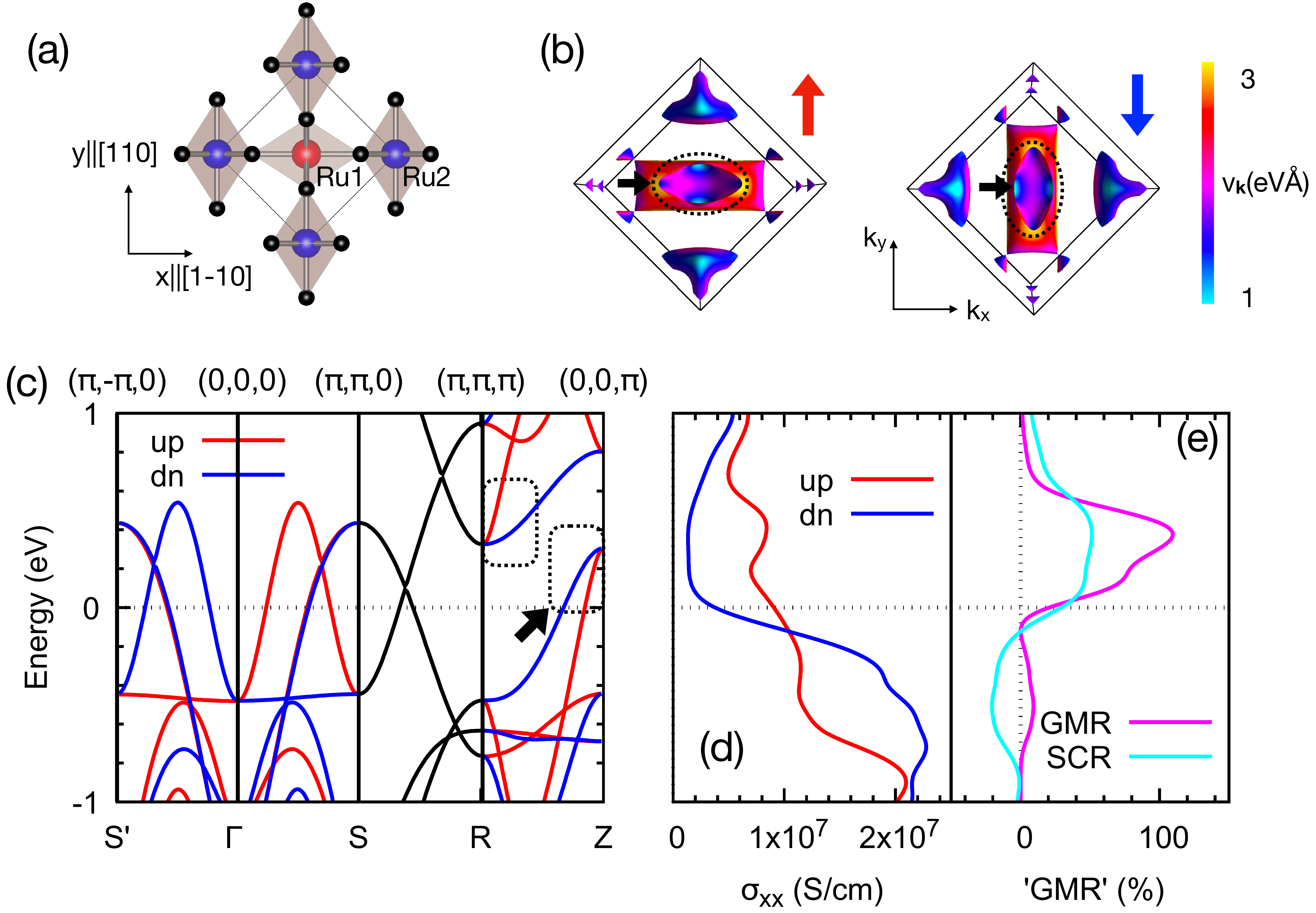} 
	\caption{\footnotesize First-principles calculations of GMR in collinear antiferromagnet RuO$_2$ \cite{Smejkal2020}. 
(a) Crystal structure of antiferromagnetic RuO$_2$. (b) Fermi surface plots for spin-up (left) and spin-down (right) states. The color coding corresponds to the group velocities, highlighting the strong anisotropy of the two spin-channel conductivities. (c) Spin projected non-relativistic energy bans of antiferromagnetic RuO$_2$  with marked anisotropic staggered spin-momentum interactions around {\bf R} and {\bf Z} points. (d) Longitudinal spin-up and spin-down conductivities. (e) GMR ratio. For comparison, the panel also shows the transverse spin current relative to the longitudinal charge current (SCR) calculated in Ref.~\cite{Gonzalez-Hernandez2020}.  } 
	\label{fig3}
\end{figure}

The spin resolved band structure is shown in Fig.~3a, and  the  spin dependent conductivities and GMR are plotted in Figs.~3b,c. The dashed rectangles in Fig.~3a depict the parts of the spectrum with the anisotropic staggered spin-momentum interaction, corresponding to our GMR model in Fig.~1. The anisotropic spin-dependent group velocities in this part of the spectrum are further highlighted in Fig.~3d. As expected from our model discussion, the difference between the {\em ab initio} spin-dependent conductivities and the corresponding antiferromagnetic GMR are maximized when the Fermi level crosses the regions with the anisotropic staggered spin-momentum interaction. Here the antiferromagnetic GMR reaches a 100\% scale. We note the correspondence of our antiferromagnetic GMR with the earlier evaluated transverse spin currents in RuO$_2$ which originate from the same difference between the spin-up and spin-down conductivities \cite{Gonzalez-Hernandez2020}. 

Fig.~4a shows {\em ab initio} calculations of  momentum and sublattice resolved energy bands in RuO$_2$, while Fig.~4b shows sublattice and spin resolved densities of states. When combined, these two figures can be used to estimate the antiferromagnetic TMR in RuO$_2$. For example, when focusing on energies around 0.5~eV (see dotted line in Figs.4a,b), we see from Fig.~4b that spin-up electrons are primarily on the spin-sublattice "1". In Fig.~4a,  we then see a valley between $\boldsymbol\Gamma$ and {\bf S} points associated with the sublattice "1". This valley has, therefore, a spin-up polarization. On the other hand, the  valley between $\boldsymbol\Gamma$ and {\bf S}$^\prime$ points has spin-down polarization. In this part of the spectrum, the antiferromagnet effectively acts as two intertwined half-metallic ferromagnets. For a valley conserved tunneling, TMR can be then estimated from the ratio of spin-up and spin-down densities of states projected on a sublattice. We emphasize that this approach is only applicable when the antiferromagnetic band structure has the staggered valley spin-splitting.  In agreement with our model discussions in Figs.~1 and 2, the {\em ab initio} TMR ratio is maximized at energies with the dominant contribution from the spin split valleys, where it reaches a 100\% scale. 

 \begin{figure}[t]
	\centering
	\includegraphics[width=.48\textwidth]{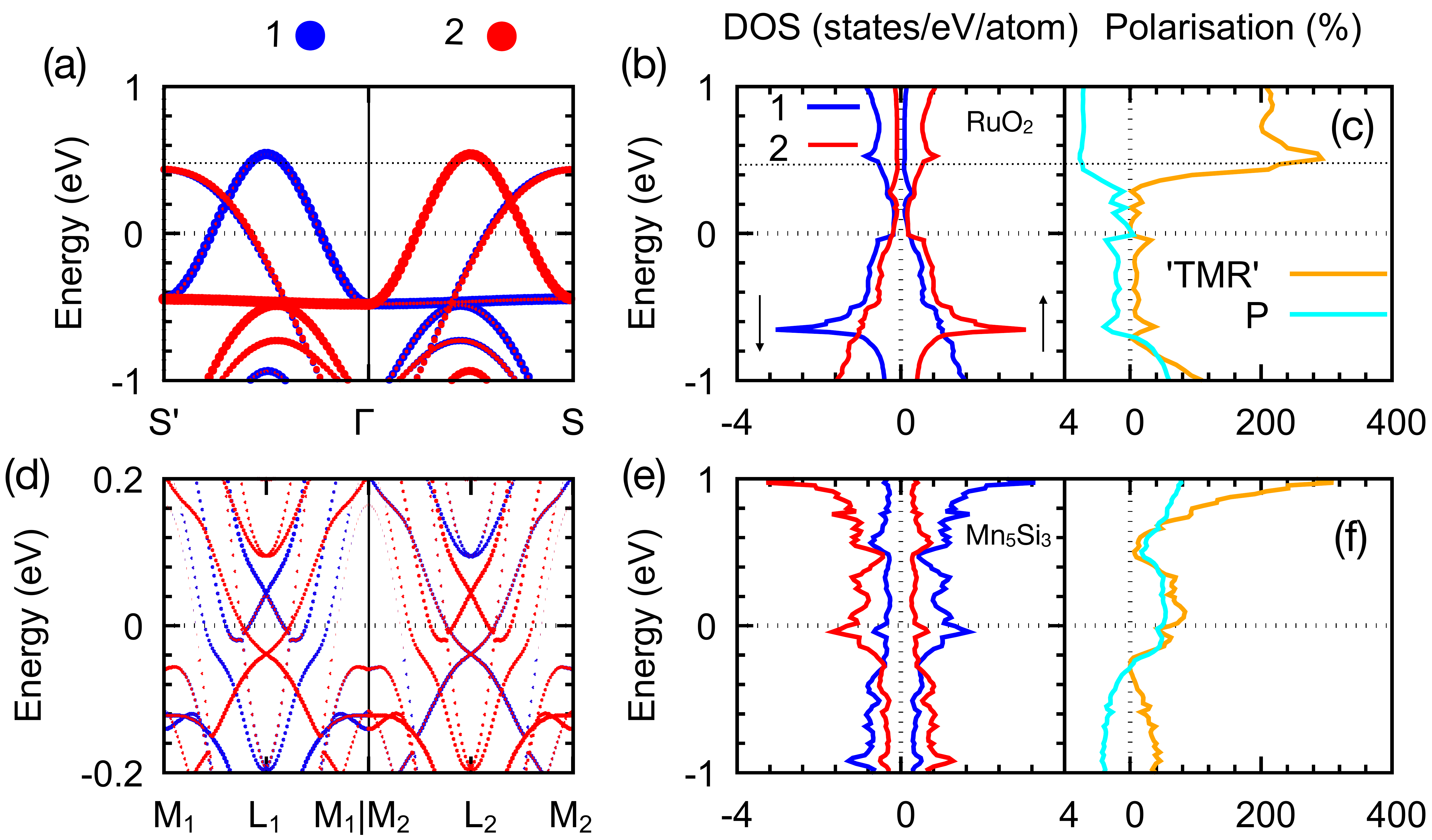} 
	\caption{\footnotesize First-principles calculations of TMR in collinear antiferromagnets RuO$_2$ \cite{Smejkal2020} and Mn$_5$Si$_3$  \cite{Reichlova2020}. (a-c) panels are for RuO$_2$, (d-f) for Mn$_5$Si$_3$. (a) Wavevector and sublattice (blue and red) resolved energy bands. (b) Sublattice (blue and red) and spin (left and right) resolved density of states as a function of energy. (c) Energy dependent TMR  and spin polarization parameter (see text). (d-f) same as (a-c) for Mn$_5$Si$_3$.} 
	\label{fig4}
\end{figure}

On a quantitative level, RuO$_2$ is not an optimal antiferromagnet for the TMR structures because of the low and weakly spin-dependent density of states at the Fermi level. A more favorable candidate is Mn$_5$Si$_3$ with a four-sublattice checkerboard collinear antiferromagnetism \cite{Reichlova2020}. The recent study in Ref.~\cite{Reichlova2020} has identified and discussed in this material the staggered spin-momentum interaction with spin-spilt valleys at time-reversal invariant momenta {\bf M}$_1$ and {\bf M}$_2$, corresponding to the model in Fig.~1c,d, as well as at {\bf L}$_1$ and {\bf L}$_2$ points. Our {\em ab initio} calculations  plotted in Figs.~4d-f show that this highly metallic antiferromagnet with staggered spin-splitting at high symmetry points in the Brillouin zone is a favorable candidate for large TMR ratios even at the Fermi level. 

For illustration, we plot in Figs.~4c,f also the spin-polarization parameter given by the relative difference of the sublattice-projected spin-up and spin-down densities of states. We see that its energy dependence correlates with the dependence of the TMR ratio. We again emphasize that TMR can be related to the  spin-polarization parameter only in the antiferromagnets with the valley-dependent spin-momentum interaction.

In the final paragraphs we comment on our anisotropic and valley-dependent models of antiferromagnets with the staggered spin-momentum interaction in the context of the excitation of the antiferromagnet by STT. In ferromagnetic stacks, in the limit of long carrier spin life-time, injected carriers with spin-polarization {\bf p} from one ferromagnet precess around the magnetization {\bf M} of the other ferromagnet. The resulting non-equilibrium spin plarization in the second ferromagnet, ${\bf s}\sim{\bf M}\times{\bf p}$, depends on magnetization {\bf M}. The corresponding (anti)damping-like STT, ${\bf T}\sim {\bf M}\times({\bf M}\times{\bf p})$,  can compete with the Gilbert damping, and thus excite  the second ferromagnet \cite{Ralph2008}. 

Earlier studies have already demonstrated that antiferromagnets, including the ones with spin-degenerate bands, can be also efficiently excited by a spin polarized current injected into the antiferromagnet \cite{Gomonay2014,Zelezny2014,Jungwirth2016}.  Here local non-equilibrium spin polarizations driving the (anti)damping-like STT at sublattice "1" and "2", ${\bf s}_{\rm 1}\sim {\bf M}_{\rm 1}\times{\bf p}$ and ${\bf s}_{\rm 2}\sim {\bf M}_{\rm 2}\times{\bf p}$, have an opposite sign on the two  sublattices since ${\bf M}_{\rm 1}=-{\bf M}_{\rm 2}$. This makes the (anti)damping-like STT due to the injected spin-current principally equally efficient in antiferromagnets as  in ferromagnets \cite{Gomonay2014,Zelezny2014,Jungwirth2016}. 

Previously, the considered spin-current injectors into the antiferromagnet were either ferromagnetic or relativistic \cite{Gomonay2014,Zelezny2014,Jungwirth2016}.  Our present study implies that antiferromagnets with the non-relativistic staggered spin-momentum interaction allow for the spin-current injection from an antiferromagnet into the other antiferromagnet.  As in the case of GMR and TMR, the anisotropic  staggered spin-momentum interaction is more favorable for antiferromagnetic STT in metallic stacks, while the valley-dependent interaction is more suitable for STT in antiferromagnetic tunneling structures.

{\em Note added.} After finishing this work we became aware of a related study by Shao et al. \cite{Shao2021} on TMR in an antiferromagnetic tunnel junction.

\section*{Acknowledgments}
We acknowledge funding form the Czech Science Foundation Grant No. 19-18623X and 21-28876J,  the Ministry of Education of the
Czech Republic Grants No. LM2018110, and LNSM-LNSpin, and  the EU FET Open RIA Grant No. 766566, Deutsche Forschungsgemeinschaft Grant TRR 173 268565370 (project A03),   Johannes Gutenberg University Grant TopDyn, the computing time granted on the supercomputer Mogon at Johannes Gutenberg University Mainz (hpc.uni-mainz.de).


%

\end{document}